\documentclass[a4paper,11pt,leqno]{article}
\pdfoutput=1 

\usepackage[english]{babel}
\usepackage[utf8]{inputenc}
\usepackage[top=1in, bottom=1in, left=0.75in, right=0.75in]{geometry}
\usepackage{amsmath}
\usepackage{natbib}
\usepackage{zref-xr}
\usepackage{zref-user}
\usepackage{hyperref}
\usepackage{graphicx}
\usepackage{caption}
\usepackage{subcaption}
\usepackage{tikz}
\usepackage[T1]{fontenc}
\usepackage{parskip} 
\usepackage{lmodern}
\usepackage{float}
\usepackage{array}
\usepackage{placeins}
\usepackage{booktabs}
\usepackage{diagbox}

\usepackage{titlesec}
\usepackage{authblk}
\usepackage{relsize}
\usepackage{microtype}
\usepackage{url}
\usepackage{lineno}
\usepackage{bm}

\usepackage{amssymb}

\titleformat{\paragraph}
{\normalfont\normalsize\bfseries}{\theparagraph}{1em}{}
\titlespacing*{\paragraph}
{0pt}{3.25ex plus 1ex minus .2ex}{1.5ex plus .2ex}
\title{\textbf{Webpage Views as a Proxy for Angler Pressure
and Effort: Insights from Bayesian Networks}}

\author[1]{Azar Taheri Tayebi\thanks{Corresponding author (email: \href{mailto:ataheritayebi@brocku.ca}{\nolinkurl{at22gg@brocku.ca}})}}
\author[2]{Julia S. Schmid}
\author[3]{Sean Simmons}
\author[2]{Mark S. Poesch}
\author[2,4]{Mark A. Lewis}
\author[1]{and Pouria Ramazi}

\affil[1]{Department of Mathematics and Statistics, Brock University, St. Catharines, Ontario, Canada}
\affil[2]{Department of Mathematical and Statistical Sciences, University of Alberta, Edmonton, Alberta, Canada}
\affil[3]{Angler's Atlas, Goldstream Publishing, Prince George, British Columbia, Canada}
\affil[4]{Department of Biological Sciences, University of Alberta, Edmonton, Alberta, Canada}
\date{}
\begin{document}
\maketitle

\section{Abstract}
 
\par Reliable angler activity data inform fisheries management. Traditionally, such data are gathered through surveys, but an innovative cost-effective approach involves utilizing online platforms and smartphone applications. 
These citizen-sourced data were reported to correlate with conventional survey information. However, the nature of this correlation--whether direct or mediated by intermediate variables--remains unclear. 
We applied BNs to data from conventional surveys, the Angler's Atlas website, the MyCatch smartphone application, and environmental data across Alberta and Ontario, Canada, to detect probabilistic dependencies. 
Using Bayesian model averaging, we quantified the strength of connections between variables. 
Waterbody webpage views were directly related to daily and weekly-aggregated boat counts in Ontario (51\% and 100\% probability) and to weekly-aggregated creel survey-reported fishing duration in Alberta (100\%).
This highlights the value of citizen-sourced data in providing unique insights beyond meteorological factors, with online interest serving as a potentially reliable proxy for angler pressure and effort.\\
\noindent\textbf{Keywords}: Angler activity, Bayesian network, citizen-sourced data, citizen-reported data, conventional surveys, machine learning
\section{Introduction}
Recreational fishing is a popular activity across the globe, with more than 10\% of people participating and catching billions of fish every year, contributing economic and social benefits \citep{arlinghaus2009recreational, kelleher2012hidden, hyder2018recreational}. 
Many recreational fisheries are facing overfishing, mortality, and fish diseases \citep{post2002canada,cooke2004role, lewin2006documented}. 
Comprehensive assessment data on resources (e.g., fish population and species) and on fishery (e.g., effort and selectivity) may enable recreational fisheries management to prevent crises through measures like size and bag regulations 
 \citep{post2008angler,venturelli2017angler}.
 
\par Data can be collected through conventional surveys such as creel surveys and aerial surveys \citep{eero2015does}.  
Creel surveys are conducted by direct interactions with anglers, either on-site while they are fishing or at the conclusion of their fishing trips \citep{murphy1996fisheries}. The collected data is limited in time and space as the collection can be expensive and time-consuming \citep{van2015imputing,volstad2006comparing}. 
Nevertheless, these conventional surveys are often considered the most reliable sources of data on angler activity, providing information that is generally assumed to represent reality.

Consequently, cost-effective citizen-sourced data became popular \citep{fischer2023boosting,gutowsky2013smartphones}. 
Mobile applications (apps) are modern sources of citizen-reported data. A survey on experts in the recreational fishery field showed that in most countries, angler app data usage is likely to increase significantly over the next 5--10 years \citep{skov2021expert}.  
Websites also serve as an additional tool for gathering data on angler activity \citep{martin2012using}. 
Anglers utilize these platforms to seek real-time updates on fishing conditions and to share information about their fishing expeditions \citep{sims2017characterizing}. 

\par Nevertheless, citizen-sourced data do not readily represent that of conventional surveys. 
Angler apps used for collecting fishing data may be subject to biases due to non-random participation. 
Who uses these apps can be influenced by factors such as smartphone ownership and demographics \citep{venturelli2017angler}. 
This could result in data skewed towards certain groups \citep{jiorle2017determining}. 
There may even be intentional data manipulation because some anglers wish to influence regulations \citep{sullivan2003exaggeration}. 
The design of the app, location of users, and internet connectivity can also bias the data \citep{papenfuss2015smartphones,jiorle2017determining}.

\par  Despite these limitations, several studies demonstrated the potential of citizen-sourced data to serve as a supplementary or alternative source for conventional survey data, particularly with respect to catch rates and angler pressure (i.e., the intensity of fishing activity, often measured by total fishing effort, number of anglers, or hours fished; Table {\ref{similarresearch}}) \citep{jivthesh2022comprehensive, johnston2022comparative, papenfuss2015smartphones}. 
For example, the iNaturalist website and app provided comprehensive data with a high level of completeness compared to other reporting methods such as telephone, email, and mixed-mode approaches \citep{taylor2022modern}. 
Data from iAngler app and creel surveys conducted by the Marine Recreational Information Program (MRIP) showed minimal differences, with discrepancies exhibiting an almost zero central tendency \citep{jiorle2016assessing}.
Likewise, mean catch rates reported via Fangstjournalen app platform were similar to those obtained through conventional surveys on a Danish island \citep{gundelund2021evaluation}. 

\par However, the nature of the dependence between citizen-sourced and conventional surveys remains unclear.
 Specifically, if the relationship is indirect and mediated by some intermediate variables, then the use of citizen-sourced data to predict conventional survey outcomes would be limited, if not zero. 
 For example, if citizen-reported catch rates are indirectly related to creel-based catch rates through the intermediate variable temperature (e.g., cold weather discourages angling and reporting), knowing the temperature alone would suffice to predict creel-based catch rates--citizen-reported catch rates provide no further information. 

 \par This defined our research goal: Is there a direct or indirect relationship between citizen-sourced and conventional survey-based angler activity measured by catch rates, fishing duration, number of trips, and number of boats? 
 To answer this question, we employed Bayesian networks (BNs) \citep{koller2009probabilistic}, a type of probabilistic graphical model that allows the detection of dependencies between variables.
 Our case study was the Lower Bow River and Upper Oldman River within three fisheries management zones in Alberta, Canada, in 2018, as well as 98 randomly selected lakes across 15 management zones in Ontario, Canada, in 2018 and 2019. 
 The target species included brook trout, walleye, and lake trout. 
 Webpage views from the Angler's Atlas website and fishing trips and catches reported by anglers through the Angler's Atlas website and MyCatch app in Alberta and Ontario were used as citizen-sourced data. 
 Conventional survey-reported datasets included creel surveys in Alberta and aerial surveys in Ontario. The environmental variables for both provinces included air temperature, total precipitation, wind speed, relative humidity, solar radiation, and degree days for the waterbodies.
\begin{table}[p]
\caption{Examples of previous research on comparing citizen-sourced data to conventional surveys on angler activity.}
\centering
 \includegraphics[width=0.93 \linewidth, height=0.9\textheight  ]{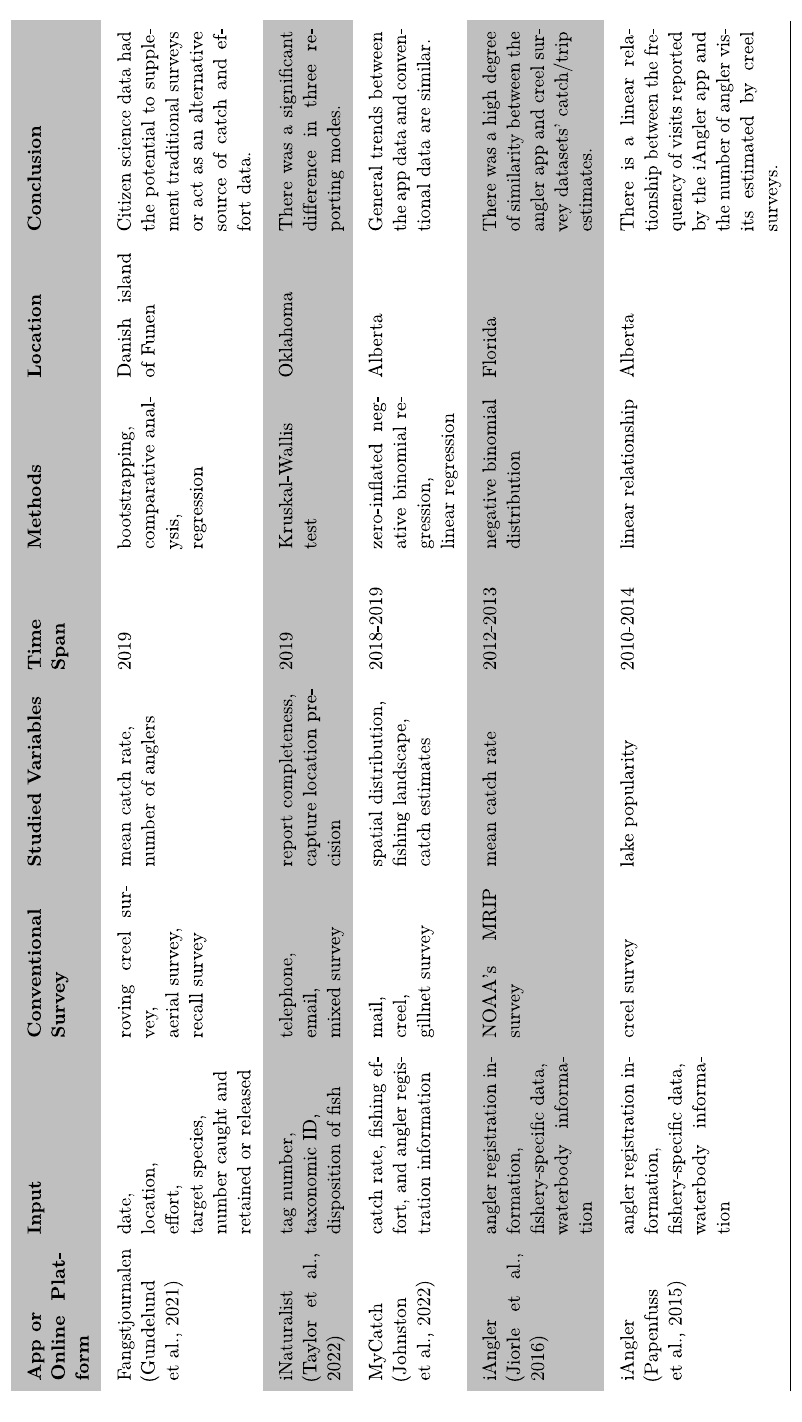}
\label{similarresearch}
 \end{table}
 
\begin{figure}
    \centering
    \includegraphics[width=1 \linewidth , height=0.5\textheight ]{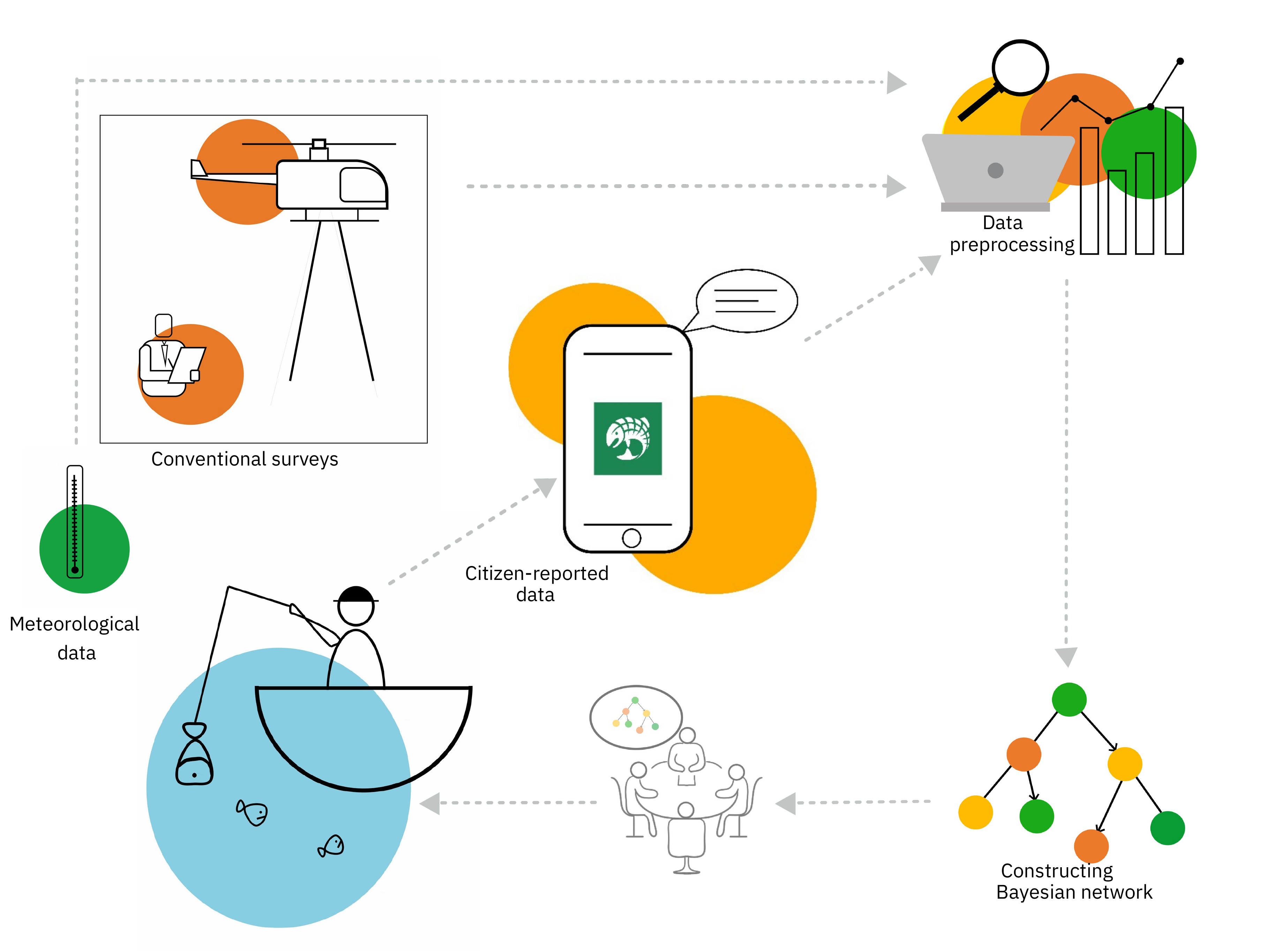}
    \caption{\small A conceptual illustration of the methodology used in this paper, that is the process of collecting data through conventional surveys, smartphone applications, and online platforms, followed by the construction of a Bayesian network to aid fishery management in decision-making. 
    An angler utilizes online platforms to find information about waterbodies and report fishing experiences and uses a smartphone app to record fishing trip data.
    Creel surveys or aerial surveys gather information about the angler trips. 
    The data is then used to construct the Bayesian network structure to reveal the conditional dependencies between the conventional survey and citizen-sourced variables in the presence of meteorological variables at as potential confounders. 
    Fishery managers may leverage the insights derived from this network structure to implement relevant actions, such as stocking and regulations. The management decision-making did not take place in this project and was not part of the methodology.}
    \label{Illustration}
\end{figure}
\section{Materials and Methods}
First, we discretized the data collected from the mentioned variables in Alberta and Ontario. Second, we learned a BN for each dataset using the Bayesian Information Criterion (BIC) score (Figure \ref{Illustration}) and analyzed the direct and indirect relationships between the variables within the BNs.
Finally, we employed Bayesian model averaging (BMA) to quantify the uncertainty associated with the estimated connections between the variables.


\subsection{Data}
The Alberta case study comprised the lower section of the Bow River and the upper section of the Oldman River and its major tributaries in southwestern Alberta, from June to October 2018. The Ontario case study included 98 lakes in Ontario. The data sources included the Angler's Atlas website and correspondent mobile app MyCatch, two creel surveys, an aerial survey, and a weather simulation model, described below.\\
\noindent\textbf{Citizen-sourced data}
\par Citizen-sourced data were collected by the Angler's Atlas website and MyCatch mobile app. The Angler's Atlas platform, established in 1999 by Goldstream Publishing Incorporated based in Prince George, British Columbia, Canada, provides detailed information on 330,000 waterbodies across Canada through its website (\url{www.anglersatlas.com}). 
Each waterbody has its own webpage featuring details such as location, popular species, and markers for hotspots and boat launches. 
The webpages can be viewed by anyone and do not require a membership on the platform. 
Anglers can use these webpages to gather trip-planning information or to report details about their fishing experiences. 
Variable ``webpage views'' was defined as the number of unique visits on a day.
Several webpage views by the same individual were counted as one view.
``Webpage views'' can be considered as an indicator of angler pressure, because visiting a specific waterbody's webpage reflects people's interest in obtaining information about the waterbody and planning trips \citep{xiang2015adapting}.
Webpage visits in the last few days (seven in our case) were considered since trips are typically planned a few days in advance \citep{nyblom2014making}.

\par On April 1, 2018, the MyCatch mobile app was launched by the Angler's Atlas team and app data collection began on May 11, 2018, through email campaigns targeting the subscribers of Angler's Atlas in Alberta \citep{johnston2022comparative}. 
Through both the Angler's Atlas online platform and the MyCatch app, anglers could report details about their fishing trips, such as date, location, duration, and number of fish caught by species. Anglers could also report trips with no catches. 
The primary caught species reported in Alberta and Ontario were brook trout, walleye, and lake trout. 
The data were anonymized by daily aggregation and included only completed trips, i.e., those with fully reported information on the trip. 
When a trip was reported for a specific waterbody, it indicated that the trip took place anywhere on that waterbody.
We refer to the collected data as ``citizen-sourced data,'' with no distinction between the app and online platform. 
 
\par
\begin{figure}[htbp] 
  \centering
  
  \begin{subfigure}[b]{0.7\linewidth} 
      \begin{tikzpicture}
          \node[anchor=south west, inner sep=0] (image) at (0,0) 
              {\includegraphics[width=\linewidth, height=0.35\textheight]{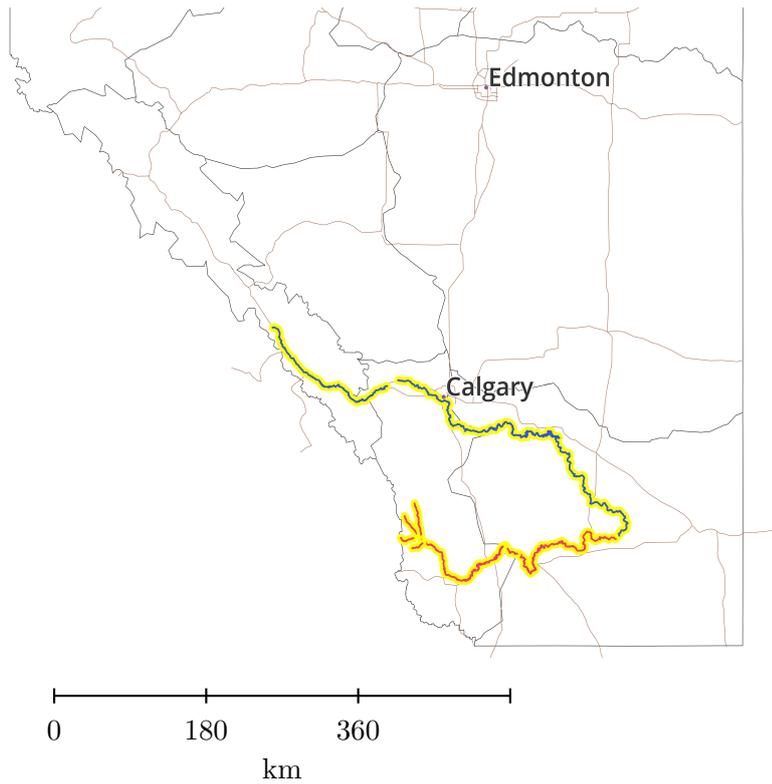}};
          
          \draw[thick] (1,-0.5) -- (7,-0.5); 
          \foreach \x in {1,3,5,7} {
              \draw[thick] (\x,-0.6) -- (\x,-0.4); 
          }
          \node[below] at (1,-0.7) {0};
          \node[below] at (3,-0.7) {180};
          \node[below] at (5,-0.7) {360};
          \node[below] at (4,-1.2) {km}; 
      \end{tikzpicture}
      \caption{Alberta study area with creel survey locations.}
      \label{alberta_map_with_scale}
  \end{subfigure}
  
  \vspace{0.8cm} 
  
  \begin{subfigure}[b]{0.7\linewidth} 
      \begin{tikzpicture}
          \node[anchor=south west, inner sep=0] (image) at (0,0) 
              {\includegraphics[width=\linewidth, height=0.35\textheight]{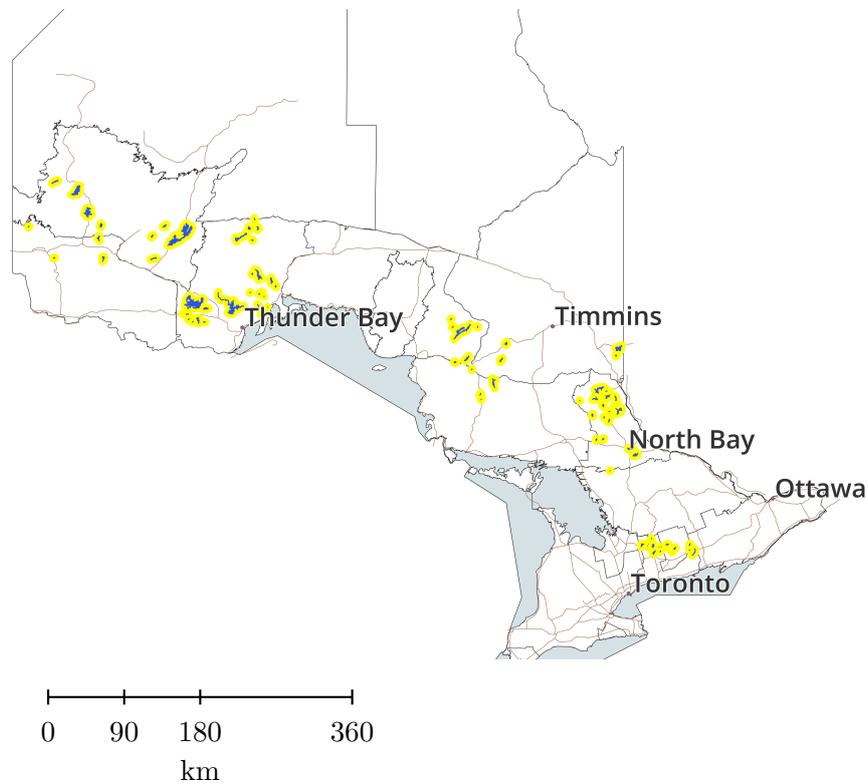}};
          
          \draw[thick] (1,-0.5) -- (5,-0.5); 
          \foreach \x in {1,2,3,5} {
              \draw[thick] (\x,-0.6) -- (\x,-0.4); 
          }
          \node[below] at (1,-0.7) {0};
          \node[below] at (2,-0.7) {90};
          \node[below] at (3,-0.7) {180};
          \node[below] at (5,-0.7) {360};
          \node[below] at (3,-1.2) {km}; 
      \end{tikzpicture}
      \caption{Ontario study area with aerial survey locations.}
      \label{ontario_map_with_scale}
  \end{subfigure}
  
  \caption{(a) Alberta study area, including the Lower Bow River and Upper Oldman River and its principal tributaries, where creel surveys were conducted in 2018. (b) Ontario study area, including 98 lakes where aerial surveys were conducted in 2018-2019.}
  \label{map}
\end{figure}

 We used citizen-sourced data from the entire Bow River and Oldman River systems, which span all three fisheries management zones in Alberta. 
There were 233 reports of fishing trips from these two river systems between May 11 and October 31, 2018. 
For Ontario, a total of 8,621 fishing trips were reported over 1,670 unique waterbodies and 310 unique days between May 19, 2018 and August 29, 2019 (Figure \ref{map}).

\par For each waterbody and day \textit{i}, the data included the total number of trips, denoted $N_i$, and for each trip $j$ the number of fish caught $C_{ij}$ and the time spent fishing $D_{ij}$.
We computed the total fishing duration  $D^{\textit{total}}_{i}$ [hour], total number of fish caught $C^{\textit{total}}_{i} $ [number of fish], and mean catch rate $CPUE_{i}$ (``catch per unit effort'') [fish/hour], for each day \textit{i}, as follows:
\begin{equation}
 \label{eq1}
D^{\textit{total}}_{i} = \sum\limits_{j=1}^{N_i} D_{ij},
\end{equation}
\begin{equation}
 \label{eq3}
 C^{\textit{total}}_{i} = \sum\limits_{j=1}^{N_i} C_{ij},
\end{equation}
\begin{equation}
\label{eq6}
 CPUE_{i} =  \frac{ C^{\textit{total}}_{i}}{D^{\textit{total}}_{i}}.
\end{equation}
For simplicity, we refer to app-reported $D^{\textit{total}}_{i}$ as ``app fishing duration'' and  $CPUE_{i}$ as ``app catch rate.''
These formed the two app variables used in this study.
The other citizen-sourced variable was ``webpage views'' presenting the number of times a waterbody's webpages was visited in the last seven days.  
\\
\noindent\textbf{Creel surveys}
\par Data from two creel surveys in Alberta in 2018 were used \citep{johnston2022comparative}.
 The first survey was conducted by Alberta Environment and Parks (AEP) along the approximately 100 kilometers of Lower Bow River, from June through the end of November 2018 \citep{christensensurvey}. The second creel survey was carried out by the Alberta Conservation Association (ACA) along a stretch of 199 kilometers encompassing the Upper Oldman River and its principal tributaries, the Livingstone River, Dutch Creek, and Racehorse Creek, from June to October 2018 \citep{hurkett2019angler}. 
We used creel data on both rivers which were collected from angler interviews during roving surveys at the public access points  \citep{hurkett2019angler, ripley2006bow, malvestuto1996sampling}. 
The surveys were conducted across spatial and temporal strata. 
Spatially, this contained four reaches on the Bow River and three on the Oldman River system. 
Temporally, this happened during shifts in the morning (8:00--15:00) or evening (15:00--22:00)--although the exact hour and date of each interview was recorded.
During interviews, 2751 anglers from the Bow River (across 122 days) and 963 anglers from the Oldman River (across 88 days) were requested to provide information about their fishing trips, including the date, time duration of the trip, and quantity of fish caught. Angler interviews included both complete and incomplete trip data.  
We excluded two samples due to daily fishing durations exceeding 24 hours.

\par We obtained the following creel-survey variables in the same way as with the corresponding app variables, using Equations \ref{eq1} to \ref{eq6}: ``total fishing duration of anglers sampled on day $i$'' $D^{\textit{total, sampled}}_{i}$, ``number of fish caught by anglers sampled on day $i$'' $C^{\textit{total, sampled}}_{i}$, and ``catch per unit effort of anglers sampled on day $i$''  $ CPUE_{i}$. 
As creel surveys were not designed to measure total fishing duration, we instead used the mean value of the variable, denoted $D^{\textit{mean}}_{i}$ [hour/trip] on day $i$ as follows: 
\begin{equation}
\label{eq5}
D^{\textit{mean}}_{i} = \frac{D^{\textit{total, sampled}}_{i}}{N_i^{sampled}}.
\end{equation}
where $N_i^{sampled}$ is the total number of trips reported by anglers sampled on day $i$.
For simplicity, we refer to the creel survey-reported $D^{\textit{mean}}_{i}$ as ``creel fishing duration'' and $ CPUE_{i}$ as ``creel catch rate'', which formed the two survey variables in the Alberta dataset.\\
\noindent\textbf{Aerial surveys}
    \par We utilized data from Ontario's inland lake ecosystem collected through the Broad-scale Monitoring (BsM) program. 
    The dataset included 98 randomly selected lakes over 50 ha, stratified by spatial (15 fisheries management zones), temporal (weekend/holiday and weekday), and lake-size categories. 
    Data were gathered during 101 aerial surveys conducted between 9:00 and 17:00 over 78 randomly selected days from June to August in 2018 and 2019 (BsM cycle 3). 
    The variable ``number of boats,'' was defined as the number of actively fishing boats observed during the surveys, serving as an indicator of angler pressure and the only survey variable in the Ontario dataset. \\
\noindent\textbf{Meteorological data} 
 \par Daily averages of meteorological data, considered as potential intermediate variables, were obtained using the BioSIM tool \citep{regniere1996landscape}. 
 BioSIM is a software developed by Natural Resources Canada that runs weather-driven simulation models using geographic weather data. 
 Based on the four nearest weather stations within a radius of 300 km from the centroid of the waterbody, the software adjusted the data for elevation, latitude, and longitude differences, and restored variability to long-term averages. 
 Historical daily weather observations were utilized, and a bi-linear interpolation method was applied \citep{regniere2017biosim}. 
 We used the software to obtain data for the waterbodies, focusing on variables known to be related to or influencing angler's behavior, such as the daily air temperature \citep{gundelund2022investigating, kendall2021winds}, precipitation \citep{shaw2021angler}, wind speed \citep{gundelund2022investigating, agmour2020impact}, relative humidity \citep{shaw2021angler}, solar radiation \citep{ shaw2021angler, cooke2017fishing, speers2012catch}, and degree days \citep{ speers2012catch}.
\par A temporal binary day-type variable ``is weekend'' was used to indicate whether the fishing trip occurred on a weekday or a weekend \citep{kendall2021winds}. \\
\noindent\textbf{The Alberta and Ontario datasets} 
\par We constructed two final (daily-aggregated) datasets: the Alberta dataset and the Ontario dataset (Table \ref{table:variables}). 
The Alberta dataset was created by combining the creel survey data and app data for the two river systems on days when creel data were available.
Each sample of the dataset included six angler activity variables: ``app catch rate,'' ``app fishing duration,'' ``app number of trips,'' ``webpage views,''  ``creel catch rate,'' and ``creel fishing duration,'' along with the seven auxiliary variables described earlier. 
The dataset had a total of 185 daily samples: 101 from Bow River and 84 from Oldman River, and an overall of 75 reported fishing trips from citizen-sourced data.
There were initially 233 reported trips in Alberta.
However, during the merging of the app data with the creel data, zero values were assigned to
app variables that did not have a value at the specified date and waterbody in the creel dataset. These
zero assignments to citizen-sourced variables were not data imputation. The reason why there was no
data for a certain waterbody or day was that no one reported on that day and for that waterbody. Thus,
by definition, the corresponding app variables would be zero.
In the same way, the Ontario dataset was formed by combining aerial survey data with app data for the waterbodies and dates with available aerial data, resulting in 1,128 daily samples over 98 waterbodies and 18 citizen-sourced reported trips. 
Each sample included five angler activity variables: ``app catch rate,'' ``app fishing duration,'' ``app number of trips,'' ``webpage views,'' and ``number of boats,'' and the same seven auxiliary variables used in the Alberta data (Table \ref{table:mytable1}; Supplementary Material Figures S1, S2, and S3).\\
\noindent\textbf{Discretization}
\par Continuous variables were discretized because the applied package for the BN learning algorithm required discrete data. 
Non-binary variables were discretized into three bins using the equal quantile method, ensuring that each bin contained approximately an equal number of samples \citep{sun2007using} (Supplementary Material, Figures S4, S5, and S6).  
Consequently, each variable had the statuses ``low,'' ``medium,'' and ``high'' (Figures S4, S5, and S6). 
Discretization into three bins was shown to be optimal in previous applications of BNs in ecological systems \citep{milns2010revealing, yu2004advances}. 
The discretization was performed separately for the Alberta and Ontario datasets.
\par If the majority of the values of a variable were zero, the first bin included only the zero values, and the remaining values were discretized into two bins with an equal number of samples. 
In the Alberta dataset, this was the case with the variables ``app number of trips'' (59\%), ``app fishing duration'' (59\% zeros), and ``app catch rate'' (71\% zeros). 
In the Ontario dataset, this applied to the variables ``number of boats'' (58\% zeros), ``app catch rate'' (99\% zeros), ``app number of trips'' (98\% zeros), ``app fishing duration'' (98\% zeros), ``webpage views'' (75\% zeros), and ``total precipitation'' (57\% zeros).

\par Due to the high proportion of zero values in the datasets, we conducted an additional experiment on two weekly-aggregated datasets for waterbodies in Alberta and Ontario. 
Each sample was assigned the calendar week number based on its calendar date. 
Then citizen-sourced and survey variables were aggregated on a weekly basis, e.g., ``number of boats'' would indicate the total number of boats surveyed during the calendar week. 
Meteorological variables were averaged over the days of the week for which any of the survey-reported variables were available, and ``is weekend'' was removed. 
In Ontario, although the aggregation reduced the dataset to 320 samples, it decreased the percentage of zeros in the app-reported and survey variables to a maximum of 18\% and 12\%, respectively. 
For data discretization, the quantile method was applied to all variables except for ``total precipitation,'' which had a high proportion of zeros (43\%). 
For this variable, the first bin was designated to represent zero values, while the nonzero values were divided into two additional bins (bins 2 and 3) using the quantile method. 
In Alberta, the aggregated data included 66 weekly samples, with a maximum of 44\% zeros in app variables and 35\% in ``webpage views.'' 
For data discretization, the quantile method was performed for all variables, except for ``app catch rate'' (44\% zeros values), ``app fishing duration'' (42\% zeros), ``app number of trips'' (42\% zeros), and ``webpage views'' (35\% zeros). 
Similarly, for these variables, the first bin was allocated to represent zero values, and nonzero values were divided into two bins using the quantile method.

\begin{table}
\centering
\caption{\textscale{}{List of variables included the Alberta and Ontario datasets.}}
\label{table:mytable1}

\includegraphics[width=0.95\linewidth]{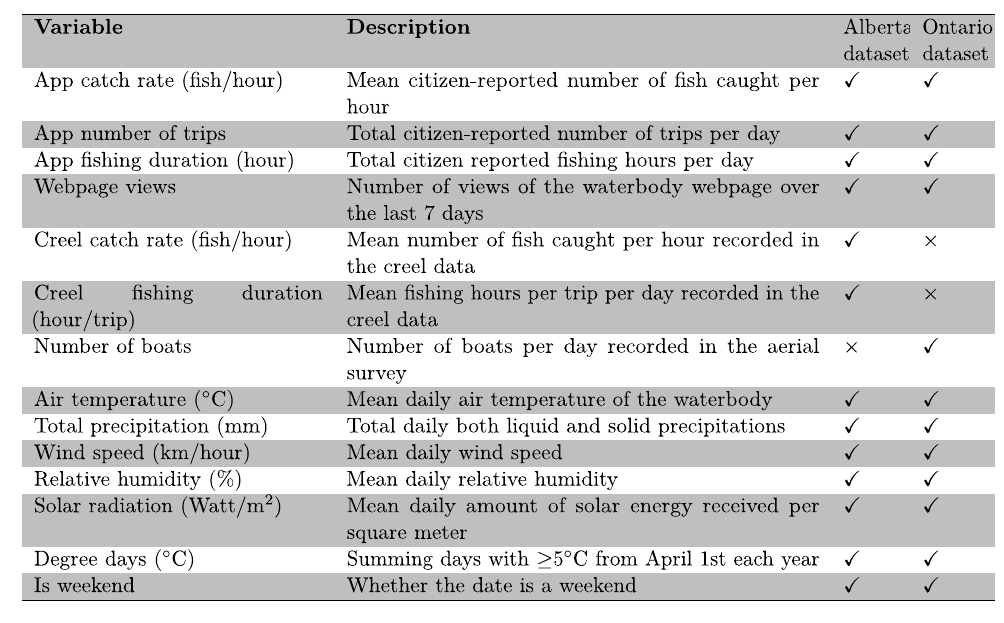}

\end{table}

\begin{table}
    \centering
    \caption{Summary of raw (original), daily-aggregated, and weekly-aggregated survey-based/citizen-sourced/combined datasets, highlighting the mean of non-zero values and the count of zeros for conventional survey-based and citizen-sourced variables.}
    \includegraphics[width=0.95\linewidth]{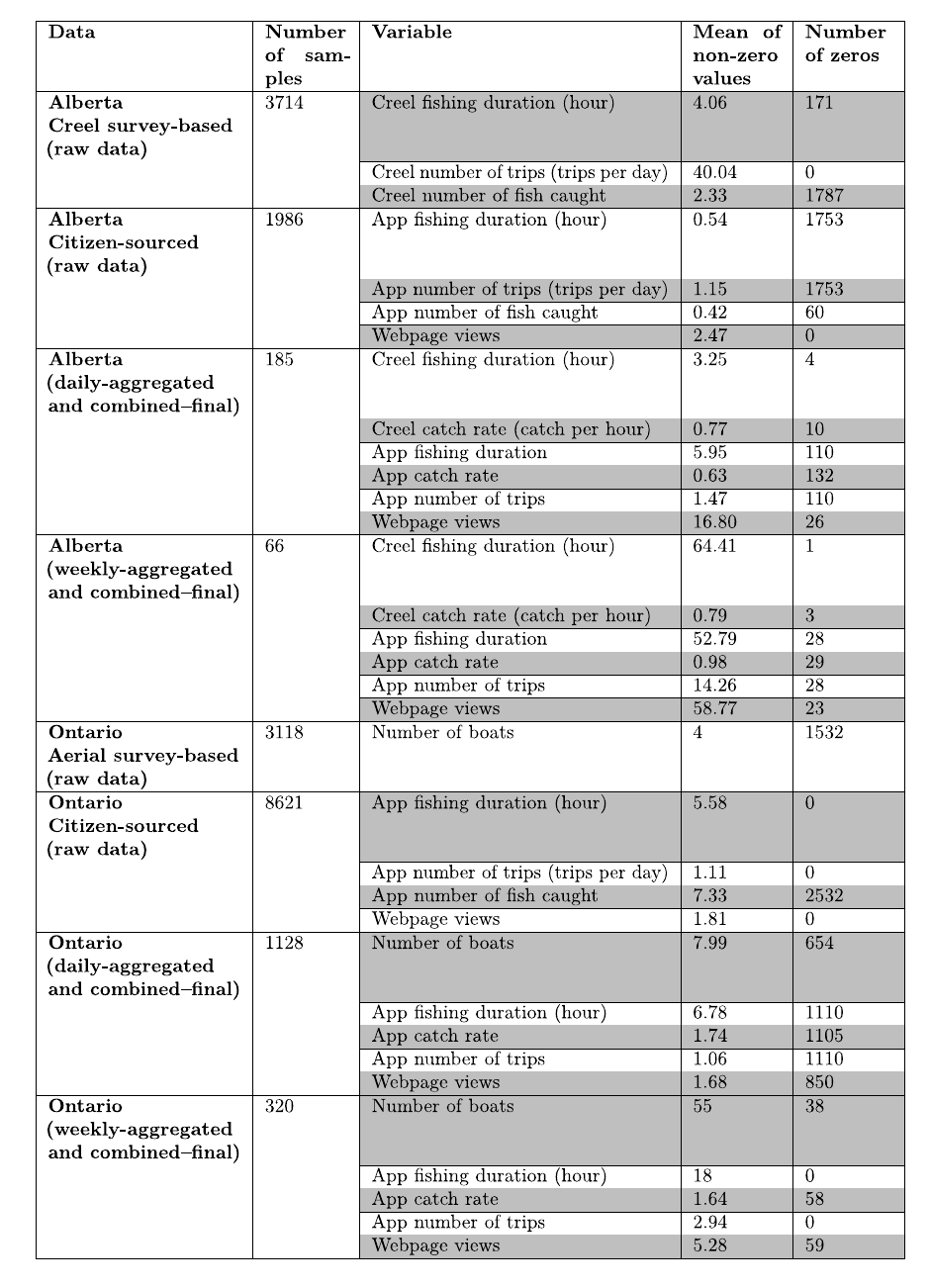}
    
    \label{table:variables}
\end{table}


\subsection{Bayesian network structure and analysis}
While the concept of direct and indirect relatedness can be challenging to investigate, a directed acyclic graph (DAG) can be used to make definitions precise and undertake analysis \citep{ramazi2022myxobolus}. 
In a causal model that is represented by a DAG, the parents of each node are considered as the \textit{direct} causes of that variable and the ancestors as the \textit{indirect} causes. More generally and out of the framework of causality, one can define direct and indirect \textit{relationships} without DAGs by using the notion of \textit{probabilistic conditional independence}. Given a specified set of random variables, two variables $X$ and $Y$ are said to be directly related if they neither are marginally independent nor become independent conditioned on any subset of the other variables. 
In contrast, variables $X$ and $Y$ are indirectly related if they are dependent but become independent once conditioned on a subset of the other variables, which we refer to as \emph{intermediate variables}. 
Variables $X$ and $Y$ are not (probabilistically) related if they are marginally/statistically independent \citep{nadkarni2001bayesian,eden1992analysis}.
\par
\begin{figure} 
    \centering
    \includegraphics[scale=0.73]{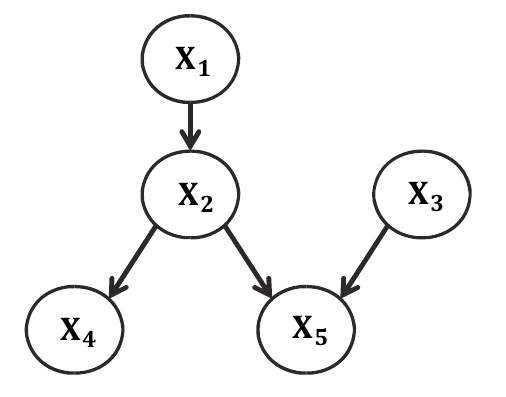}
    \caption{\small An example of a Bayesian network structure representing the direct and indirect relationships between five variables.
    Node \(X_1\) is directly connected to node \(X_2\) and indirectly connected to nodes \(X_4\) and \(X_5\) through node \(X_2\). 
    If node \(X_2\) is observed, \(X_1\) becomes d-separated from variables \(X_4\) and \(X_5\). 
    This means that \(X_1\) is independent of \(X_4\) and \(X_5\) given \(X_2\), i.e., \(X_1 \perp X_4, X_5 \mid X_2\). 
    The observed node \(X_2\) also d-separates \(X_4\) from \(X_5\). 
    Additionally, \(X_1\) and \(X_3\) are d-separated if \(X_5\) is not observed, but they become dependent if $X_5$ is observed.
}
    \label{examplegraph}
\end{figure}

 A BN is a tool to identify conditional independencies among specified variables. A BN consists of a \emph{structure} that is a DAG over the variables considered as nodes and \emph{parameters} that are the conditional probability distributions of each variable conditioned on its parents in the DAG \citep{koller2009probabilistic,ramazi2023bayesian}. 
In the structure, an edge between two variables implies a direct relationship: a dependence that does not turn into independence regardless of what other variables are observed (conditioned on). 
Conditional independence in a BN is captured by the notion of \textit{d-separation} (directed separation) in the network structure. 
A collider on a path is the subgraph $A\to B\leftarrow C$ on the path, that is, a triple of nodes $A$, $B$, and $C$ where $A$ and $C$ are linked to $B$. The node $B$ is called the collider node. 
A path between nodes $X$ and $Y$ is active, given observed variables $\bm{Z}$ if two conditions are met: (i) for every collider on this path, either the collider node itself or at least one of its descendants (children, grandchildren, etc.) are observed, and (ii) no other node on this path is observed. 
Two variables $X$ and $Y$ are d-separated given observed variables $\bm{Z}$ if there is no active path between $X$ and $Y$ in the network. 
Two variables $X$ and $Y$ being d-separated (given variable(s) $\bm{Z}$) in a BN implies that they are independent given $\bm{Z}$, i.e., $X \perp Y \mid\bm{Z}$, where $\bm{Z}$ can be the empty set (Figure \ref{examplegraph}). 
Moreover, under the commonly made assumptions of faithfulness, all (conditional) independencies between the variables are also captured by the d-separations in the BN.

\par The BN structure readily reveals the direct and indirect relationships between the variables. The structure of the network can be obtained by searching over the space of possible DAGs and selecting the DAG with the maximum BIC (or other) score(s)
indicating the simplest DAG that best fits the data \citep{pearl1988probabilistic}. 
BNs have a foundation in ecological applications \citep{heggerud2024predicting, johnson2012integrated,milns2010revealing}. 
An introduction to applying and interpreting BNs in ecological datasets is given in \citep{ramazi2021exploiting}.

  We obtained two BNs, one for the Alberta data, and one for the Ontario data using the command \texttt{learn.network} in the \texttt{bnstruc} package in \texttt{R} \citep{franzin2017bnstruct}, which applied the Silander-Myllymaki algorithm to find the globally optimal BN structure \citep{silander2012simple} in terms of the BIC score. This score-based algorithm searches for the structure with the highest BIC by optimizing small BN sub-structures and then utilizing the outcomes to optimize larger structures, continuing this process until finding the optimal global BN structure. The output graphs from these algorithms are often not fully directed because there can be more than one DAG resulting in the same BIC score. 
  This implies that flipping the direction of the undirected links in either way will yield the same BIC score, as long as it does not result in a directed cycle or a new v-structure, that is, a triple of nodes $X$, $Y$, $Z$, where $X$ and $Y$ are linked to $Z$ and $X$ and $Y$ are not connected. 
  Therefore, each of the graphs obtained by an arbitrary orientation of the undirected links can be considered as the ``best'' BN. 
  Once the BN structure was obtained, we investigated the direct or indirect nature of the relationships between the angler activity variables using d-separation. 
  The directions of the arrows may not be interpreted causally \citep{ramazi2021exploiting}, unless further assumptions are made.\\
\noindent\textbf{Bayesian Model Averaging}
\par When only a single BN is used as the final model, equal trust is assigned to all links, and any uncertainty about the structure selection is disregarded \citep{ wintle2003use}. 
Bayesian model averaging (BMA) is a solution to consider the uncertainty in the dominated network by taking the expectation of the quantity of interest across all possible network structures \citep{broom2012model, hoeting1999bayesian, hinne2020conceptual}. 
For example, instead of checking whether variables $X$ and $Y$ are linked in the best BN structure, we find the probability of the two being linked across all possible BN structures. 
In the process, probabilities for the existence are given to the links, whereby links that are observed in the network structures with higher scores during the search receive higher probabilities \citep{liao2023improved, hwang2005bayesian, tian2012bayesian, madigan1994model}. 
 The number of possible BN structures in a global search is super-exponential \citep{tian2012bayesian}. 
 To reduce the computational complexity, we estimated the probability by considering only a portion of the networks (Eq. \ref{eq4}).
 Thus, a probability of 100\% for a link would indicate its presence in only the sampled structures and not necessarily all structures.
 The sampled structures were chosen based on the
 Markov Chain Monte Carlo (MCMC) \citep{brooks1998markov, friedman2003being} method. 
 The MCMC algorithm involved defining a Markov Chain over the space of possible network structures, generating a set of structures through a random walk in the chain and using a scoring criterion to evaluate and probabilistically accept or reject changes to the network structure, such as adding, removing, or reversing edges \citep{madigan1994model}. 
 The expectation (probability) of two nodes $X$ and $Y$ being adjacent (denoted $X$-$Y$) in the ``best'' BN equals
\begin{equation}
\label{eq4} 
\begin{aligned}
 E(X - Y )& = \sum_{\hat{ \mathcal{G}_i}} P(\hat{ \mathcal{G}_i}|D)f(\hat{ \mathcal{G}}_i) \\ &= \sum_{\hat{ \mathcal{G}_i}: X - Y \in \hat{ \mathcal{G}}_i} P(\hat{ \mathcal{G}_i}|D),
\end{aligned}
\end{equation}
where  $\hat{ \mathcal{G}_i}$ is a DAG sampled by MCMC, $D$ is the dataset, $P(\hat{ \mathcal{G}_i}|D) $ is the likelihood of the sampled DAG,
\begin{equation*}
f(\hat{\mathcal{G}}_i) = 
\begin{cases} 
1, & \text{if } X \to Y \in \hat{\mathcal{G}}_i 
\text{ or } X \leftarrow Y \in \hat{\mathcal{G}}_i \\
0, & \text{otherwise,}
\end{cases}
\end{equation*}
and where the summation goes over all sampled DAGs $\hat{ \mathcal{G}_i}$ \citep{koivisto2004exact}. 
To obtain the probability of existing links between the variables using the MCMC algorithm, we utilized the function \texttt{plinks(bdgraph.obj)} in the \texttt{BDgraph} package version 2.72 in \texttt{R} \citep{mohammadi2015bdgraph}.  The package would find the probability of undirected links only, so the probabilities that we reported were for both directions, that is the probability of a link $X$→$Y$ or $Y$→$X$ exists between $X$ and $Y$.
 
 
 \section{Results}
\begin{figure}
\begin{subfigure}[b]{0.70\linewidth} 
\centering
\includegraphics[width=1\linewidth,height=0.4\textheight]{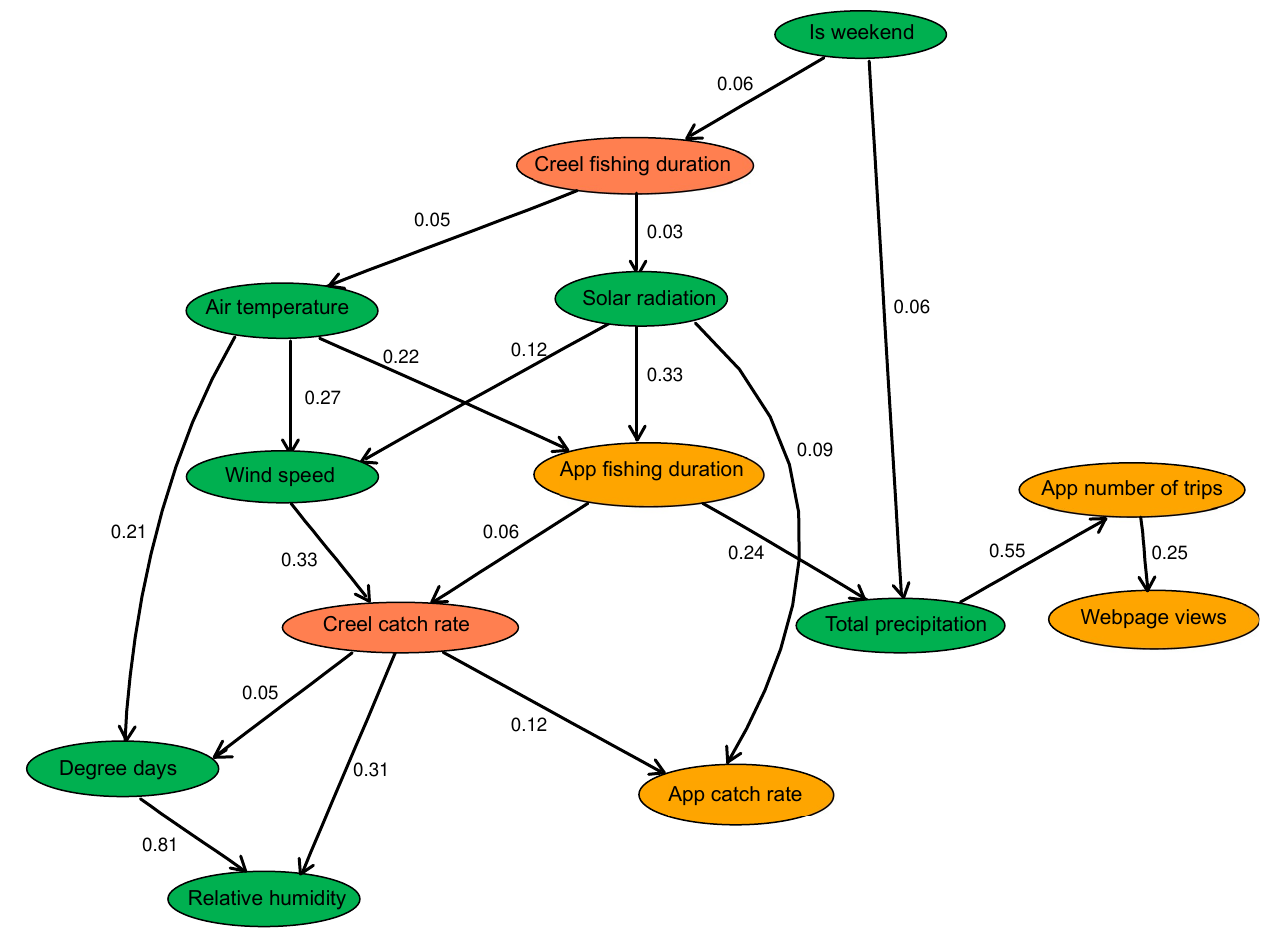}
\caption{\textscale{0.9}{Daily-aggregated Alberta dataset}}
\end{subfigure}
\\
\vspace{1pt}
   \begin{subfigure}[b]{0.8\linewidth} 
  \centering
    \includegraphics[width=1\linewidth , height=0.43\textheight]{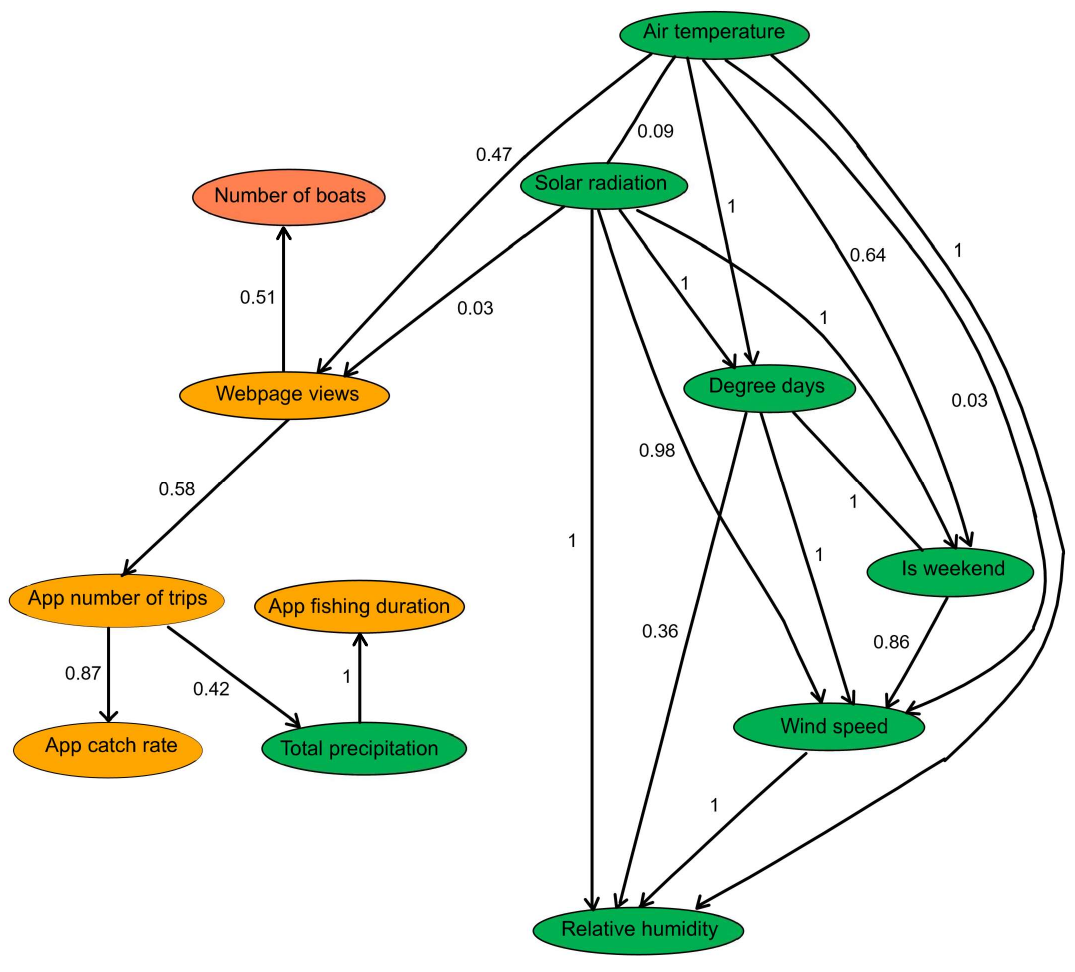}
    \caption{\textscale{0.9}{Daily-aggregated Ontario dataset}}
  \end{subfigure}
  \caption{\textscale{0.8}{Bayesian network structures from the datasets of (a) Alberta (185 daily samples) and (b) Ontario (1128 daily samples). Green nodes signify environmental and time variables, and yellow and coral nodes represent citizen-sourced and conventional survey variables, respectively. 
  Edges indicate probabilistic dependencies between variables. The number on each is the probability of that single edge existing in the ``true'' network. 
  An edge without direction can be directed in both ways as its direction has no impact on the BIC score, as long as directed cycles and new v-structures are not created.}}
  \label{BN_Silander}
\end{figure}

\begin{figure} 
  \begin{subfigure}[b]{0.6\linewidth} 
   \centering
    \includegraphics[width=1\linewidth , height=0.27\textheight]{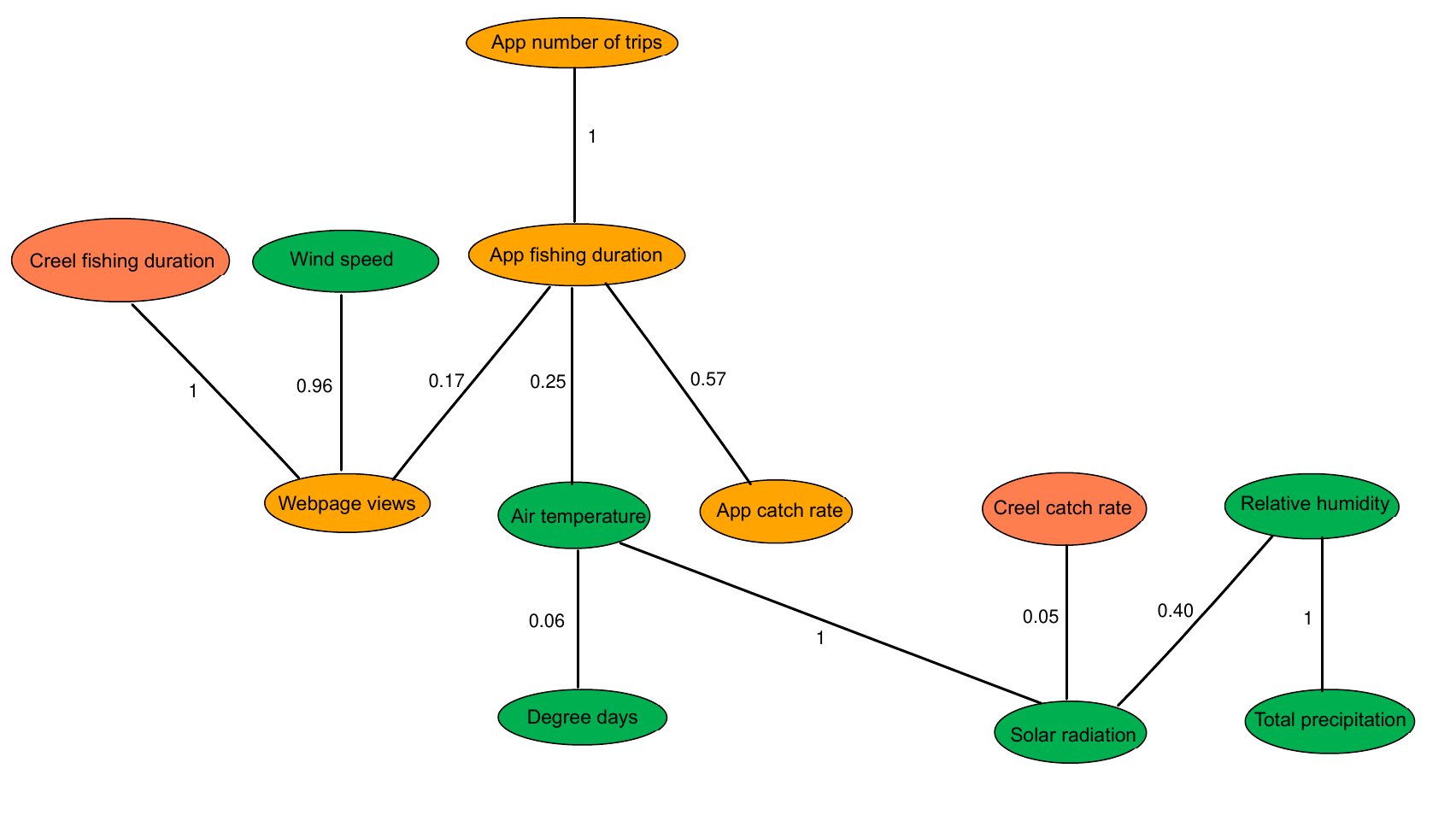}
    \caption{\textscale{0.9}{Weekly-aggregated Alberta dataset}}
  \end{subfigure}\\
 \vspace{1pt}
   \begin{subfigure}[b]{0.8\linewidth} 
  \centering
    \includegraphics[width=1\linewidth , height=0.45\textheight]{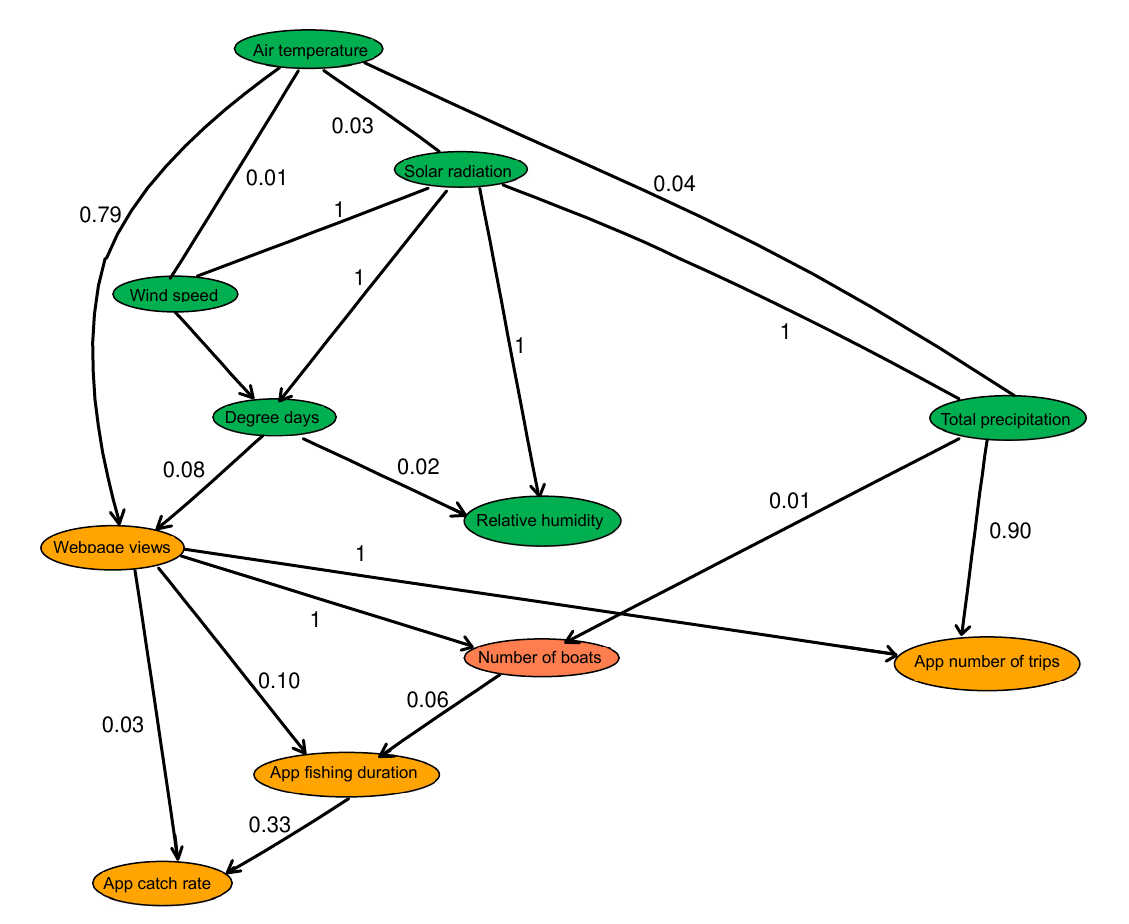}
    \caption{\textscale{0.9}{Weekly-aggregated Ontario dataset}} 
  \end{subfigure}
    \caption{ \textscale{0.8}{ The Bayesian networks constructed from the weekly-aggregated (a) Ontario and (b) Alberta datasets, with 320 and 66 weekly samples, respectively. 
    Green nodes signify environmental and yellow and coral
nodes represent citizen-sourced and survey variables, respectively. 
Edges indicate probabilistic dependencies
between variables. 
The number on each edge is the probability of that single edge existing in the ``true'' network. 
}}
    \label{weekly_BN_1}
\end{figure}

In the BN structure of the daily-aggregated Alberta dataset (Figure \ref{BN_Silander}a), ``creel catch rate'' was directly related to ``app catch rate'' (edge probability 12\%)  and ``app fishing duration,'' (6\%) and indirectly related to ``app number of trips,'' via the intermediate variables ``total precipitation'' and ``app fishing duration.'' 
``Creel catch rate'' was also indirectly related to ``webpage views,'' via ``app fishing duration,'' ``total precipitation,'' and ``app number of trips.''  
``Creel fishing duration'' was indirectly related to  ``app fishing duration,'' being d-separated given intermediate variables ``solar radiation,'' and ``air temperature.'' 
It was also indirectly related to ``app catch rate'' via the weak connection to ``solar radiation'' (3\%). 
In the weekly-aggregated Alberta dataset (Figure \ref{weekly_BN_1}a), ``creel fishing duration'' was directly connected to ``webpage views''(100\%) and d-separated from all other variables given ``webpage views.''
``Creel catch rate'' was only connected to ``solar radiation'' (5\%).

\par In the BN structure of the daily-aggregated Ontario dataset (Figure \ref{BN_Silander}b), the aerial survey-reported ``number of boats'' was directly related to ``webpage views'' (51\%).
Given ``webpage views,'' ``number of boats'' was d-separated from all other variables. 
In the weekly-aggregated Ontario dataset, ``number of boats'' was directly linked to ``webpage views'' (100\%) and ``app fishing duration'' (6\%; Figure \ref{weekly_BN_1}b).   
 \section{Discussion}
We used BNs to analyze the relationship between angler activities from conventional surveys and citizen-sourced data and meteorological variables in Alberta and Ontario. 
``Webpage views'' was linked directly to aerial angler pressure in daily and weekly-aggregated Ontario data and to ``creel fishing duration''--an indicator of angler effort--in weekly-aggregated Alberta data, suggesting that online interest may serve as a strong indicator of the intensity of angler activity (i.e., pressure and/or effort). 
In daily-aggregated Alberta, a direct but weaker connection was observed between survey-based and app-based catch rates. 
These direct connections suggest that citizen-sourced data provide unique information on creel survey variables, beyond meteorological influences.

\par 
To our knowledge, this is the first study to propose ``webpage views'' as a low-cost proxy for angler activity, reflecting anglers' tendency of checking conditions online before visiting waterbodies. 
However, without direct evidence that webpage viewers are anglers or how they use these platforms, this remains a hypothesis. 
The absence of a direct link between ``webpage views'' and creel-based variables in daily-aggregated Alberta may be due to the earlier MyCatch app adoption in 2018, making anglers there more reliant on the app compared to Ontario anglers in 2018–2019, who likely used the Angler’s Atlas website for trip planning. 
Future work could explore visits to other webpages or search engine trends \citep{martin2012using} as predictors of angler pressure or effort.
\par Air temperature was connected directly to ``app fishing duration'' in Alberta and indirectly to ``number of boats'' in Ontario, aligning with studies on the correlation between angler activities and air temperature \citep{midway2023heat, gundelund2022investigating, sylvia2020influence}.
Precipitation was strongly connected to app-based angler pressure in both provinces--except weekly Alberta--supporting the idea that changing precipitation may reduce daily trips \citep{hunt2006potential}. 
Its indirect link to catch rate aligns in part with evidence that precipitation shapes seasonal catch and fish dynamics \citep{baptista2016influence, guy1991seasonal}.
Daily Alberta data showed a weak creel-citizen catch rate link, absent weekly, possibly contrasting prior close ties \citep{johnston2022comparative, gundelund2021evaluation, jiorle2016assessing}. 
Despite low-probability links, app variables might still predict daily creel rates, warranting further investigation.
Moreover, wind speed was linked to creel-based catch rate, consistent with research indicating that wind-induced turbulence hinders boat fishing \citep{speers2012catch, zischke2012catch, agmour2020impact}.

\par
The high proportion of zeros in the daily datasets, especially in Ontario, reflects the inherent characteristics of recreational fishing behavior and data collection methods.
For app-reported variables such as ``app number of trips'' and ``app fishing duration,'' zeros arise on days when no fishing activity was reported on the MyCatch app, which may occur due to unfavorable weather conditions, low angler participation, or non-use of the app. 
These zeros are not data gaps but carry meaningful information about the conditions under which fishing activity or reporting does not occur. 
Retaining these zeros in the analysis allowed us to uncover the relationships between environmental and temporal factors and the absence of reporting of fishing activity, offering insights into the broader dynamics of angler activity.  
Weekly aggregation obscured these day-to-day dynamics while highlighting broader patterns, showing variables that predict weekly but not daily trends, and vice versa.

\par Analyzing a more extensive dataset from Ontario compared to Alberta, the Bayesian model averaging indicated a higher probability for the detected direct relationships between the variables. 
Namely, in the Ontario dataset, there was a great match between the results from the model averaging, which accounted for multiple plausible structures, and the global search, which identified a single best-fitting structure.
This reinforces the credibility of the identified relationships and their relevance for understanding and managing ecological systems \citep{lavine2019frequentist, ellison2004bayesian}.
However, due to computational complexity, Bayesian model averaging was performed on a sampled subset of BN structures.

\par Differences in the connections between the Alberta and Ontario datasets could result from, in addition to the difference in the data sizes, ecological and social factors and differences in how the variables were defined (e.g., mean or total fishing durations). 
Firstly, the structure of a BN depends on the probabilistic dependencies between variables, which can change with the presence or absence of certain variables \citep{kitson2024impact}. 
For example, ``wind speed'' and ``degree days'' were indirectly connected through intermediate variable ``creel catch rate'' in the Alberta dataset, but the two variables were directly connected in the Ontario dataset where ``creel catch rate'' was absent. 
The differences could also stem from unobserved intermediate variables. Environmental factors such as water temperature and air pressure, socioeconomic factors such as angler income, and social events occurring in the waterbodies can shape angler activity \citep{meyer2023effects, dabrowksa2017understanding, cantrell2004recreational}. 

\par Secondly, differences in survey methods could introduce bias. 
Creel surveys recorded fishing durations of incomplete fishing trips, likely leading to underestimation \citep{eckelbecker2019comparing}. Conversely, app-reported fishing durations corresponded to completed trips, but it is unclear if anglers reported the time actively spent fishing only or the entire trip duration, possibly causing overestimation. 
Additionally, weather variables were averaged for entire river systems or sections, but weather conditions can vary significantly at individual waterbodies, which may influence angler activities \citep{dippold2020spatial}.

Thirdly, spatial variability within and between datasets can influence angler activities and, consequently, the BN structures. For example, the Bow River is dominated by brown trout, while the Oldman River is home to cutthroat trout \citep{johnston2022comparative}.  
Anglers may be drawn to specific fish species and genus, affecting catch rates and pressure \citep{hunt2019catch, hunt2002understanding}.
Other spatial factors, such as water quality, aesthetics, privacy, and accessibility impact angler choices \citep{hunt2005recreational, moeller1972fishermen, hunt2006potential}. 
Local income levels, fishing skills, and motivation of anglers further influence behavior, but these factors are often difficult to measure \citep{gundelund2021changes, hunt2019catch}.

\par  This study has several limitations. 
First, it focuses on only two provinces. 
Evaluating the methodology in other regions or with broader datasets would likely improve its generalizability. 
Second, more detailed spatial, temporal, and demographic data (e.g., waterbody size, water temperature, population, and income) could further clarify the relationships between conventional surveys and citizen-sourced data. 
Third, the high proportion of zeros in the citizen-sourced dataset—particularly in Ontario—suggests low app participation, probably due to limited promotion following its launch. 
Because the app was introduced earlier and more extensively in Alberta, this discrepancy is understandable. 
A higher percentage of reporting anglers may uncover relationships to conventionally surveyed variables.
Increasing engagement and accurate data collection will require user-friendly interfaces, interactive features, well-designed prompts, and training for anglers \citet{venturelli2017angler}.


\section*{Dataset availability}
The dataset used in this study is available in \citep{dataset1}.

\section*{Acknowledgements}
This project, including the use of the dataset, was reviewed and approved by the Research Ethics Board of the Alberta Research Information Services (ARISE, University of Alberta), study ID MS5 Pro00102610.
We acknowledge the support of the Government of Canada’s New Frontiers in Research Fund (NFRF), NFRFR-2021-00265.
We acknowledge Joel Knudsen, Jamie Svendsen, and Clayton Green for their assistance in gathering and preparing the data.
We acknowledge Fiona D. Johnston for her valuable support with the creel data.
PR acknowledges funding from an NSERC Discovery Grant.
\section{Appendix}
\appendix

\bibliographystyle{apalike}
\bibliography{bibb}

\end{document}